\documentstyle[12pt]{article}
\begin{document}
\hfill{CCUTH-95-06}
\vskip 0.3cm
\begin{center}
{\large {\bf Resummation in Hard QCD Processes}}
\vskip 1.0cm
Hsiang-nan Li
\vskip 0.5cm
Department of Physics, National Chung-Cheng University, \par
Chia-Yi, Taiwan, Republic of China
\end{center}
\vskip 1.0cm

\vskip 1.0cm
\centerline{\bf Abstract}
\vskip 0.3cm
We show that the resummation of large radiative corrections in QCD 
processes can be performed both in covariant gauge and in axial gauge. 
We extend the resummation technique to inclusive processes,
concentrating on deeply inelastic scattering, Drell-Yan production, and 
inclusive heavy meson decays in end-point regions. Leading and 
next-to-leading logarithms produced by radiative corrections are organized 
into Sudakov factors, whose expressions are the same as those derived from
the conventional Wilson-loop approach. The gauge invariance of the 
resummation in the end-point region is demonstrated explicitly.

\newpage

\centerline{\large \bf I. INTRODUCTION}
\vskip 0.5cm

One of the essential subjects in perturbative QCD (PQCD) is to organize
large radiative corrections. For processes involving both a high and a low 
energy scale, large logarithms associated with ultraviolet, collinear and 
soft divergences are produced by loop integrals. Collinear and soft 
divergences may overlap to form double logarithms. These logarithms 
appear in products with coupling constant $\alpha_s$, and spoil the 
perturbative expansion. It has been found that radiative corrections 
in kinematic end-point regions of inclusive processes, such as deeply 
inelastic scattering, Drell-Yan production \cite{S,CT,KM} and heavy meson 
decays $B\to X_u l\nu$ and $B\to X_s\gamma$ \cite{KS,L1,LY1}, introduce
double logarithms. Double logarithms also occur in exclusive
processes, such as Landshoff scattering \cite{BS}, hadron form factors 
\cite{LS}, Compton scattering \cite{CL} and heavy-to-light transitions 
$B\to\pi(\rho)$ \cite{LY2} and $B\to D^{(*)}$ at maximal recoil \cite{L2}. 
In order to have a meaningful PQCD analysis of these processes, 
the important logarithms must be summed to all orders. 

The conventional approach to the summation of large radiative corrections 
in inclusive processes is based on the factorization of Wilson loops
\cite{GK,CC,FST}, which are formed by the classical trajectories of quarks 
involved in scattering. These Wilson loops collect important infrared 
logarithms, and are treated by renormalization group (RG) methods. An 
alternative approach is the resummation technique developed in \cite{CS}, 
which is more closely related to PQCD factorization theorems. This 
technique has been applied to exclusive processes mentioned above. Recently, 
it was extended to inclusive decays of heavy mesons \cite{L1,LY1}, and 
results are found to be the same as those from the Wilson-loop formalism.

In this paper we shall discuss the application of the resummation 
technique to inclusive processes in a more rigorous and detailed manner, 
concentrating on deeply inelastic scattering and Drell-Yan production.
The equivalence of the resummation technique and the Wilson-loop formalism
is justified by demonstrating that they lead to Sudakov factors which are 
equal up to next-to-leading logarithms. We shll show that a resummation 
can be performed both in covariant gauge $\partial \cdot A=0$ and in axial 
gauge $n\cdot A=0$, and that the resummation is independent of the 
gauge vector $n$. This is a subject which cannot be addressed in the 
Wilson-loop approach, since the vector $n$ has been fixed in the direction 
of classical trajectories of quarks. The gauge dependence of the 
resummation was not discussed either in \cite{CS}, where a vector $n$
with ${\bf n}_T=0$ was chosen.

We show how to perform a resummation in covariant 
gauge and in axial gauge in Sec. II, taking a jet subprocess as an example. 
We then apply the technique to deeply inelastic scattering in Sec. III. 
The factorization of a relevant structure function into a 
distribution function, a jet function, a soft function and a hard 
scattering amplitude is formulated. Leading and next-to-leading 
logarithms in these convolution factors are organized into a Sudakov factor. 
It is justified explicitly that the dependence of the Sudakov factor on the 
gauge vector $n$ cancels among the distribution, soft and jet functions. 
Drell-Yan production and inclusive heavy meson decays are studied in Sec. 
IV and Sec. V, respectively. We conclude that the resummation results for 
the above processes are the same as those from the Wilson-loop approach.
Sec. VI is the dicussion. Some details of calculation are presented in the
Appendix.
\vskip 2.0cm

\centerline{\large\bf II. COVARINT GAUGE AND AXIAL GAUGE}
\vskip 0.5cm

The resummation technique was originally developed in axial gauge 
$n\cdot A=0$ \cite{CS}. In this section we shall show that it can also be 
performed in covariant gauge $\partial\cdot A=0$, and leads to the same 
results. The physical meaning of the gauge vector $n$, and the concept 
of the resummation are in fact understood easilier in covariant gauge.

We take as an example a jet subprocess defined by the matrix element 
\begin{equation}
J(p,n)=\langle 0|P\exp\left[ig\int_0^\infty dz n\cdot A(nz)\right]
q(0)|p\rangle\;,
\label{j}
\end{equation}
where $q$ is a light quark with momentum $p$. The abelian case of this 
subprocess has been discussed in \cite{C}. Our formulation presented 
below is more general and also suitable for the nonabelian case. The 
path-ordered exponential in Eq.~(\ref{j}) is the consequence of the 
factorization of collinear gluons, possessing momenta parallel to $p$, 
from a full process. This factorization will become transparent in Sec. 
III. For convenience, we assume that $p$ has a large light-cone 
component $p^+$, and all its other components vanish. A general diagram 
of $J$ is shown in Fig.~1(a), where the path-ordered exponential is 
represented by an eikonal line along the vector $n$ with collinear gluons 
attaching to it. The eikonal line behaves like a scalar particle carrying 
a color source, whose propagator is written as $1/(n\cdot l)$, $l$ being 
the momentum flowing through the eikonal line. The Feynman rule for an 
eikonal vertex is given by $n_\mu T^a$, $T^a$ being a color matrix. All 
these rules can be derived from the definition of $J$ in Eq.~(\ref{j}). 
To ensure that the collected gluons lead only to the collinear divergences 
from the region with $l$ parallel to $p$, $n$ does not lie on the light 
cone.

It is easy to find that $J$ contains double logarithms from the overlap of 
collinear and soft divergences by investigating the lowest-order diagrams
in Figs.~1(b), the self-energy correction to the quark $q$, and 1(c), the 
vertex correction with a gluon connecting the quark line and the eikonal 
line. In covariant gauge both Figs.~1(b) and 1(c) produce double logarithms. 
In axial gauge $n\cdot A=0$, however, the path-ordered exponential 
reduces to an identity, and thus Fig.~1(c) does not exist. In this case $n$ 
goes into the gluon propagator, $(-i/l^2)N^{\mu\nu}$, with
\begin{equation}
N^{\mu\nu}=g^{\mu\nu}-\frac{n^\mu l^\nu+n^\nu l^\mu}
{n\cdot l}+n^2\frac{l^\mu l^\nu}{(n\cdot l)^2}\;,
\label{gp}
\end{equation}
and only Fig.~1(b) leads to double logarithms. The vanishing of Fig.~1(c)
in axial gauge can also be confirmed by the fact that the attachment of
a gluon to an eikonal vertex gives a factor $N^{\mu\nu}n_\nu=0$.

We show how to resum the double logarithms in $J$ in covariant 
gauge. The essential step in the resummation procedures is to derive a 
differential equation $p^+dJ/dp^+=CJ$ \cite{L1,LY1,LY2}, where the 
coefficient function $C$ contains only single logarithms, and can be 
treated by RG methods. Note that the path-ordered exponential is 
scale-invariant in $n$, and thus $J$ must depend on $p$ and $n$ through 
the ratio $(p\cdot n)^2/n^2$. Hence, the differential operator $d/dp^+$ 
can be replaced by $d/dn$ using a chain rule,
\begin{equation}
p^+\frac{d}{dp^+}J=-\frac{n^2}{v\cdot n}v_\alpha\frac{d}{dn_\alpha}J\;,
\label{cr}
\end{equation}
$v=(1,0,{\bf 0}_T)$ being a vector defined by $p=p^+v$.

Equation (\ref{cr}) simplifies the analysis tremendously, because $n$ 
appears only in the Feynman rules for the eikonal line, while $p$ may flow 
through the whole diagram in Fig.~1(a). The differentiation of each eikonal 
vertex and of the associated eikonal propagator with respect to $n_\alpha$, 
\begin{eqnarray}
-\frac{n^2}{v\cdot n}v_\alpha\frac{d}{dn_\alpha}\frac{n_\mu}{n\cdot l}
&=&\frac{n^2}{v\cdot n}\left(\frac{v\cdot l}{n\cdot l}n_\mu-v_\mu\right)
\frac{1}{n\cdot l}
\nonumber \\
&\equiv&\frac{{\hat n}_\mu}{n\cdot l}\;,
\label{dp}
\end{eqnarray}
leads to the derivative $p^+dJ/dp^+$ described by Fig.~2. The summation in 
Fig.~2 is performed over different attachments of the symbol $+$, which 
denotes the new vertex ${\hat n}_\mu$ defined by the second expression 
in Eq.~(\ref{dp}). 

If the loop momentum $l$ is parallel to $p$, the factor $v\cdot l$ 
vanishes, and ${\hat n}_\mu$ is proportional to $v_\mu$. When this 
${\hat n}_\mu$ is contracted with a vertex in $J$, in which all momenta 
are mainly parallel to $p$, it can be shown that the contribution to 
$p^+dJ/dp^+$ from this collinear region diminishes. Hence, the leading 
regions of $l$ are soft and hard. 

According to this observation, we investigate some two-loop examples 
exhibited in Fig.~3(a). If the loop momentum flowing through the new vertex 
is soft but the other one is not, only the first diagram is important, 
giving large single logarithms. In this soft region the subdiagram 
containing the new vertex can be factorized using the eikonal approximation 
as shown in Fig.~3(b), where the symbol $\otimes$ represents a convoluting 
relation. The subdiagram is absorbed into a function $K$, and the remainder 
assigned to $J$, both being $O(\alpha_s)$ contributions. The function $K$ 
is part of the coefficient $C$ introduced before. If both the 
loop momenta are soft, the four diagrams in Fig.~3(a) are equally 
important. The subdiagrams are factorized according to Fig.~3(c), which 
give $O(\alpha_s^2)$ contributions to $K$. The remainder is then the 
leading diagram of $J$.

If the loop momentum flowing through the new vertex is hard and the other 
one is not, only the second diagram in Fig.~3(a) is important. In this 
region the subdiagram containing the new vertex is factorized 
as shown in Fig.~3(d). The right-hand side of the dashed line is absorbed 
into a function $G$ as an $O(\alpha_s)$ contribution, which is another part 
of the coefficient $C$. The left-hand side is identified to be the
$O(\alpha_s)$ diagram of $J$. If both the loop momenta are hard, all 
the diagrams in Fig.~3(a) are absorbed into $G$, giving $O(\alpha_s^2)$ 
contributions.

Extending the above analysis to all orders, we derive the differential 
equation
\begin{equation}
p^+\frac{d}{dp^+}J=\left[K(m/\mu,\alpha_s(\mu))+G(p^+\nu/\mu,
\alpha_s(\mu))\right]J\;,
\label{de1}
\end{equation}
with the functions $K$ and $G$ described by the corresponding diagrams in
Fig.~4(a). In Eq.~(\ref{de1}) $\mu$ is a factorization scale, and the 
gauge factor $\nu=\sqrt{(v\cdot n)^2/|n^2|}$ in $G$ comes from the 
reexpression $\sqrt{(p\cdot n)^2/|n^2|}=p^+\nu$. A gluon mass $m$ is 
introduced to regulate the infrared divergences in $K$. We have made 
explicit the fact that $K$ and $G$ depend on a single infrared scale $m$ 
and a single ultraviolet scale $p^+$, respectively.

We consider the $O(\alpha_s)$ contribution to $K$ from Fig.~4(b), which is 
written as
\begin{eqnarray}
K&=&-ig^2{\cal C}_F\mu^\epsilon\int\frac{d^{4-\epsilon} l}
{(2\pi)^{4-\epsilon}}\frac{{\hat n}_\mu}{n\cdot l}
\frac{g^{\mu\nu}}{l^2-m^2}\frac{v_\nu}{v\cdot l}-\delta K
\nonumber \\
&=&-ig^2{\cal C}_F\mu^\epsilon\int\frac{d^{4-\epsilon} l}
{(2\pi)^{4-\epsilon}}\frac{n^2}{(n\cdot l)^2 (l^2-m^2)}-\delta K\;,
\label{k1} 
\end{eqnarray}
where ${\cal C}_F=4/3$ is a color factor and $\delta K$ an additive 
counterterm. The $O(\alpha_s)$ contribution to $G$ from Fig.~4(c), where 
the soft subtraction ensures a hard loop momentum flow, is given by
\begin{eqnarray}
G=-ig^2{\cal C}_F\mu^\epsilon\int\frac{d^{4-\epsilon} l}
{(2\pi)^{4-\epsilon}}\frac{{\hat n}_\mu}{n\cdot l}\frac{g^{\mu\nu}}{l^2}
\left(\frac{\not p+\not l}{(p+l)^2}\gamma_\nu-\frac{v_\nu}{v\cdot l}\right)
-\delta G\;,
\label{g1}
\end{eqnarray}
$\delta G$ being an additive counterterm. By evaluating Eqs.~(\ref{k1}) and 
(\ref{g1}) directly, one finds that $K$ and $G$ contain only single 
logarithms $\ln(m/\mu)$ and $\ln(p^+\nu/\mu)$, respectively, as claimed 
before. Organizing these single logarithms using RG methods, and then 
solving Eq.~(\ref{de1}), we resum the double logarithms in $J$. These 
procedures will be implemented in a detailed way in Sec. III.

We then review how to perform a resummation in axial gauge 
$n\cdot A=0$, and compare the outcomes with those obtained in covariant 
gauge. Since gluons are detatched from the eikonal line in axial gauge,
we concentrate on diagrams of the type Fig.~1(b). In this case the 
scale invariance of $J$ in $n$ is manifest in the gluon propagator 
$N^{\mu\nu}$. The chain rule in Eq.~(\ref{cr}) then still holds, but now
$d/dn$ operates on $N^{\mu\nu}$, giving
\begin{eqnarray}
-\frac{n^2}{v\cdot n}v_\alpha
\frac{d}{dn_\alpha}N^{\mu\nu}&=&\frac{n^2v_\alpha}{v\cdot nn\cdot l}
\left(N^{\mu\alpha}l^\nu+N^{\alpha\nu}l^\mu\right)
\nonumber \\
&\equiv& {\hat v}_\alpha
\left(N^{\mu\alpha}l^\nu+N^{\alpha\nu}l^\mu\right)\;.
\label{dgp}
\end{eqnarray}
The momentum $l^\mu$ ($l^\nu$) appearing at one end of the 
differentiated gluon line is contracted with a vertex the gluon 
attaches, and the vertex is replaced by the new one 
${\hat v}_\alpha$, defined by the second expression in Eq.~(\ref{dgp}).
The contraction of $l^\mu$ hints the application of the Ward identity.
Summing all the diagrams with different differentiated gluon propagators, 
and employing the Ward identity, one finds that the new vertex moves to 
the right most end of the $q$-quark line. We then obtain the derivative 
$p^+dJ/dp^+$ described by Fig.~5(a), where the square represents the new 
vertex ${\hat v}_\alpha$. 

We argue that the collinear region of the loop momentum $l$ is not 
important because of the factor $1/(n\cdot l)$ in ${\hat v}_\alpha$ with 
nonvanishing $n^2$. Hence, the leading regions of $l$ are soft and hard, 
in which the subdiagram containing the square can be factorized into a 
function $K$ and a function $G$, respectively, following the same reasoning 
as for Fig.~3. We then derive a differential equation similar to 
(\ref{de1}) with the general diagrams of $K$ and $G$ displayed in Fig.~5(b). 
The $O(\alpha_s)$ contributions to $\cal K$ from Fig.~5(c) and to 
$\cal G$ from Fig.~5(d) are given by
\begin{eqnarray}
K&=&ig^2{\cal C}_F\mu^\epsilon\int\frac{d^{4-\epsilon} l}
{(2\pi)^{4-\epsilon}}\frac{{\hat v}_\mu v_\nu N^{\mu\nu}}
{(v\cdot l)(l^2-m^2)}
-\delta K
\nonumber \\
&=&-ig^2{\cal C}_F\mu^\epsilon\int\frac{d^{4-\epsilon} l}
{(2\pi)^{4-\epsilon}}\frac{n^2}{(n\cdot l)^2 (l^2-m^2)}-\delta K\;,
\label{k2} \\
G&=&ig^2{\cal C}_F\mu^\epsilon\int\frac{d^{4-\epsilon} l}
{(2\pi)^{4-\epsilon}}{\hat v}_\mu\left(\frac{\not p+\not l}
{(p+l)^2}\gamma_\nu-\frac{v_\nu}{v\cdot l}\right)\frac{N^{\mu\nu}}
{l^2}-\delta G\;.
\label{g2}
\end{eqnarray}
Similarly, a gluon mass $m$ is introduced to serve as an infrared cutoff of 
the integral for $K$. The second expression in Eq.~(\ref{k2}) is derived in 
the Appendix: The first term $g^{\mu\nu}$ in $N^{\mu\nu}$ does not 
contribute. The contribution from the third term $-n^\nu l^\mu/(n\cdot l)$ 
cancels that from the fourth term $n^2l^\mu l^\nu/(n\cdot l)^2$. Hence, 
only the second term $-n^\mu l^\nu/(n\cdot l)$ remains. 

It is observed that Eq.~(\ref{k1}) in covariant gauge is equal to 
(\ref{k2}) in axial gauge. It is also straightforward to justify that 
Eq.~(\ref{g1}) is equal to (\ref{g2}). These equalities generalize to 
higher loops, and thus the differential equation (\ref{de1}) in covariant 
gauge is exactly the same as that in axial gauge. We conclude 
that the resummation can be performed in both covariant and axial gauges,
and the same results are obtained. In the following analyses of inclusive 
processes we shall work only in axial gauge.
\vskip 2.0cm

\centerline{\large \bf III. DEEPLY INELASTIC SCATTERING}
\vskip 0.5cm

In this section we discuss the resummation of large radiative corrections
in deeply inelastic scattering,
\begin{equation}
\ell(k)+h(p)\to \ell(k')+X\;,
\end{equation}
where $k$ ($k'$) is the initial (final) momentum of the lepton $\ell$, 
and $p$ is the momentum of the initial hadron $h$. We choose a 
center-of-mass frame of the incoming hadron and lepton, such that
$p$ and $k$ have the light-cone components $p=(p^+,0,{\bf 0}_T)$ and 
$k=(0,k^-,{\bf 0}_T)$ with $p^+=k^-$. The Bjorken variable is defined by
\begin{equation}
x=\frac{-q^2}{2p\cdot q}\equiv\frac{Q^2}{2p\cdot q}\;,
\label{bjo}
\end{equation}
$q=k-k'$ being the momentum tranfer. We shall derive the factorization 
formula for deeply inelastic scattering, which is appropriate in the 
end-point region $x\to 1$. Radiative corrections to the convolution factors 
involved in the factorization formula that produce double logarithms will
be organized using the resummation technique in axial gauge.
\vskip 0.5cm

\noindent
III. 1 Factorization

Fig.~6(a) describes the scattering of a valence quark in the hadron $h$ 
with the fractional momentum $\xi p+{\bf p}_T$ by an off-shell photon with 
the momentum $q$. ${\bf p}_T$ is a small transverse momentum, which will 
serve as the infrared cutoff of loop integrals below. The outgoing quark 
carries the momentum $p'=\xi p+{\bf p}_T+q=(\xi p^+-k'^+,k^--k'^-,{\bf p}_T
-{\bf k}'_T)$, where the components $k'^\mu$ satisfy the on-shell condition
$k'^2=0$. Including the requirement $p'^2\ge 0$, we have the constraint 
$x\le\xi\le 1$. 

We consider radiative corrections to Fig.~6(a) in axial gauge. Some
${\cal O}(\alpha_s)$ diagrams are displayed in Figs.~6(b)-(e). Fig.~6(b), 
the self-energy correction to the valence quark, and Fig.~6(c), the loop 
correction with a real gluon connecting the two valence quarks, contain 
both collinear divergences from the loop momentum $l$ parallel to $p$ and 
soft divergences from small $l$. Since the soft divergences cancel 
asymptotically as shown later, the double logarithmic corrections are 
mainly collinear. Therefore, Figs.~6(b) and 6(c) can be absorbed into a 
distribution function $\phi$ associated with the initial hadron.

In the end-point region with $k'^+\to p^+$ and $k'^-\to 0$ ({\it ie.}, 
$x\to \xi\to 1$), the outgoing quark, possessing a large minus component 
$p'^-\approx k^-$ but a small invariant mass $p'^2\approx (\xi-x)Q^2$, 
produces a jet of collinear particles. The self-energy ccorrection to 
the outgoing quark in Fig.~6(d) then gives both collinear divergences from 
$l$ parallel to $p'$ and soft divergences. Because of the asymptotic 
cancellation of the soft divergences, these double logarithms are mainly 
collinear, and Fig.~6(d) can be absorbed into a jet function $J$. Note that 
this jet function does not exist in the conventional factorization formulas 
for deeply inelastic scattering, which are appopriate for $x$ not close to 
1. This is the reason the logarithmic corrections are stronger in the 
end-point region. 

The absorption of the vertex-correction diagram in Fig.~6(e) is more 
delicate. Fig.~6(e) contains collinear divergences from $l$ parallel to 
$p$ and soft divergences. Moreover, it contains additional collinear 
divergences from $l$ parallel to $p'$ as $x\to 1$. We treat this diagram 
in different ways in these three leading regions. For $l$ parallel to 
$p$, we replace the outgoing quark by an eikonal line in the direction 
$n$, and factorize the gluon into $\phi$ from the full scattering 
amplitude. It is then clear that such an eikonal line, collecting collinear 
gluons, is necessary for the factorization of vertex-correction diagrams, 
and has been associated with the jet subprocess in Eq.~(\ref{j}). This 
arbitrary vector $n$ with nonvanishing $n^2$ does not introduce extra
collinear divergences, but soft divergences through the eikonal propagator 
$1/(n\cdot l)$. These soft divergences, differing from the original ones
which should be grouped by the classical trajectory of the outgoing quark 
in the direction $p'$, can be removed accorrding to Fig.~7(a). The soft 
subtraction is obtained by further replacing the valence quark by an 
eikonal line in the direction $v=p/p^+=(1,0,{\bf 0}_T)$. 

Similarly, for $l$ parallel to $p'$, we replace the valence quark by an 
eikonal line also along $n$, and absorb the gluon into $J$. 
The additional soft divergences are subtracted following Fig.~7(b), where 
$v'=(0,1,{\bf 0}_T)$ is a vector along $p'$ in the $x\to 1$ region. If $l$ 
is soft, all quarks are replaced by their classical trajectories, 
{\it ie.}, by an eikonal line along $v$ for the valence quark 
and by an eikonal line along $v'$ for the outgoing quark, as 
shown in Fig.~7(c). This diagram, together with the collinear subtractions 
which remove the divergences from $l$ parallel to $v$ and to $v'$, are 
factorized into a soft function $U$. Fortunately, in axial gauge 
$n\cdot A=0$ all the diagrams containing eikonal lines in the direction $n$ 
do not contribute, and only the first diagram in Fig.~7(c) is left. At last, 
when the loop momenta in Figs.~6(b)-(e) are hard, the gluons are 
grouped into a hard scattering amplitude $H$. 

Based on the above reasoning, we write down the factorization formula for a 
structure function $F(x,Q^2)$ associated with deeply inelastic scattering,
\begin{equation}
F(x,Q^2)=\int_x^1 d\xi \int d^2{\bf b}\phi(\xi,p,b,\mu)J(p',b,\mu)
U(b,\mu)H(\xi,Q,\mu)\;,
\label{fdis}
\end{equation}
graphically described by Fig.~8, where $b$ is the conjugate variable of 
$p_T$. The reason Eq.~(\ref{fdis}) is formulated in $b$, instead of $p_T$, 
space has been given in \cite{L1,LY2}. Briefly speaking, a formulation in 
$b$ space facilitates the factorization of higher-order corrections. 
The variable $b$ is introduced by a Fourier function $e^{i{\bf l}_T\cdot 
{\bf b}}$ associated with the integrand for a higher-order correction
whenever the loop momentum $l$ flows through $\phi$ or $J$. We have assumed 
that all the convolution factors in Eq.~(\ref{fdis}) depend on a single $b$ 
for simplicity. The relevant scales each convolution factor contains have 
been shown explicitly. $\phi$ and $J$, involving both a large scale $p^+$ 
(or $p'^-$) and a small scale $1/b$, possess double logarithms. $U$ and $H$, 
depending only on one scale $1/b$ and $Q$, respectively, contain single 
logarithms. 
\vskip 0.5cm

\noindent
III. 2 The distribution function $\phi$

We demonstrate how to resum the double logarithms in the distribution
function $\phi$. For a similar reason, $\phi$ depends on the ratio 
$(p\cdot n)^2/n^2$, and thus the chain rule in Eq.~(\ref{cr})
holds. Applying $d/dn$ to a gluon propagator and summing all the diagrams
with different differentiated gluon lines, we obtain the derivative 
$p^+d\phi/dp^+$ described by Fig.~9(a), where the coefficient 2 counts the 
outer most ends of the two valence quark lines, and the square represents 
the new vertex ${\hat v}_\alpha$ defined by Eq.~(\ref{dgp}).

Similarly, in the leading soft and ultraviolet regions the subdiagram 
containing the square is factorized into a function $K$ and a function $G$, 
respectively. The differential equation is then expressed as
\begin{equation}
p^+\frac{d}{dp^+}\phi=2\left[K(b\mu,\alpha_s(\mu))+G
(p^+\nu/\mu,\alpha_s(\mu))\right]\phi\;,
\label{dph}
\end{equation}
with the general diagrams of $K$ and $G$ shown in Fig.~9(b). We keep the 
gauge factor $\nu=\sqrt{(v\cdot n)^2/|n^2|}$ in $G$ in order to trace the 
gauge dependence of the resummation. It has been shown in 
Eq.~(\ref{dph}) that $K$ depends on a single small scale 
$1/b$ and $G$ depends on a single large scale $p^+$. 

The $O(\alpha_s)$ contributions to $K$ from Fig.~9(c) and to $G$ from 
Fig.~9(d) are given by
\begin{eqnarray}
K&=&ig^2{\cal C}_F\mu^\epsilon\int\frac{d^{4-\epsilon}l}
{(2\pi)^{4-\epsilon}}
\left[\frac{1}{l^2}+2\pi i\delta(l^2)e^{i{\bf l}_T\cdot {\bf b}}\right]
\frac{{\hat v}_\mu v_\nu}{v\cdot l}N^{\mu\nu}-\delta K\;,
\label{kph}\\
G&=&ig^2{\cal C}_F\mu^\epsilon\int\frac{d^{4-\epsilon}l}
{(2\pi)^{4-\epsilon}}{\hat v}_\mu
\left(\frac{\not p+\not l}{(p+l)^2}\gamma_\nu
-\frac{v_\nu}{v\cdot l}\right)\frac{N^{\mu\nu}}{l^2}-\delta G\;,
\label{gph}
\end{eqnarray}
$\delta K$ and $\delta G$ being the corresponding additive counterterms. 
In Eq.~(\ref{kph}) the second term, corresponding to the second diagram in 
Fig.~9(c), contains the factor $e^{i{\bf l}_T\cdot {\bf b}}$, such that $K$ 
is free of infrared poles. That is, $1/b$ serves as an infrared cutoff of 
the loop integral as stated before. Note that $K$ and thus the soft 
divergences vanish in the asymptotic region $b\to 0$. The second term in 
Eq.~(\ref{gph}) is the soft subtraction associated with $G$. 

A straightforward calculation of Eqs.~(\ref{kph}) and (\ref{gph}) in the 
modified minimal subtraction $({\overline{\rm MS}})$ scheme leads to
\begin{eqnarray}
K(b\mu,\alpha_s)&=&-\frac{\alpha_s}{\pi}{\cal C}_F
\ln(b\mu)
\;,\label{ok} \\
G(p^+\nu/\mu,\alpha_s)&=&-\frac{\alpha_s}{\pi}{\cal C}_F
\ln\frac{p^+\nu}{\mu}
\;,
\label{oj}
\end{eqnarray}
and $\delta K=-\delta G$. Details of the calculation refer to the 
Appendix. To perform the integral for $G$, we have chosen a space-like 
gauge vector $n$ with $n^2<0$, which avoids imaginary contributions. 
In Eqs.~(\ref{ok}) and (\ref{oj}) we keep only the
large logarithmic terms, and neglect other constants of
order unity. This is reasonable, because the factorization of $K$ and
$G$ in Eq.~(\ref{dph}) holds only in the leading regions
characterized by large logarithms. It is found that $G$ depends on the 
large scale $p^+$ through the ratio $(p\cdot n)^2/n^2$ as expected. 

Since $K$ and $G$ contain only single soft and ultraviolet logarithms, 
respectively, they can be treated by RG methods:
\begin{equation}
\mu\frac{d}{d\mu}K=-\lambda_K=
-\mu\frac{d}{d\mu}G\;.
\label{kg}
\end{equation}
The anomalous dimension of $K$, $\lambda_K=\mu d\delta K/d\mu$,
is given, up to two loops, by \cite{BS,LS}
\begin{equation}
\lambda_K=\frac{\alpha_s}{\pi}{\cal C}_F+\left(\frac{\alpha_s}{\pi}
\right)^2{\cal C}_F\left[{\cal C}_A\left(\frac{67}{36}
-\frac{\pi^{2}}{12}\right)-\frac{5}{18}n_{f}\right]\;,
\label{lk}
\end{equation}
with $n_{f}$ the number of quark flavors, and ${\cal C}_A=3$ a color 
factor. In solving Eq.~(\ref{kg}), we allow the scale $\mu$ to evolve to 
the infrared cutoff $1/b$ in $K$ and to $p^+$ in $G$. The RG solution of 
$K+G$ is then written as
\begin{eqnarray}
& &K(b\mu,\alpha_s(\mu))+G(p^+\nu/\mu,\alpha_s(\mu))
\nonumber \\
& &\hskip 1.0cm =K(1,\alpha_s(1/b))+G(\nu,\alpha_s(p^+))-
\int_{1/b}^{p^+}\frac{d{\bar\mu}}{\bar\mu}
\lambda_K(\alpha_s({\bar\mu}))\;,
\nonumber \\
& &\hskip 1.0cm =-\frac{\alpha_s(p^+)}{\pi}{\cal C}_F\ln \nu-
\int_{1/b}^{p^+}\frac{d{\bar\mu}}{\bar\mu}\lambda_K(\alpha_s({\bar\mu}))\;.
\label{skg}
\end{eqnarray}

Substituting Eq.~(\ref{skg}) into (\ref{dph}), we derive
\begin{eqnarray}
\phi(\xi,p,b,\mu)&=&
\exp\left[-2\int_{1/b}^{Q}\frac{d p^+}{p^+}
\left(\int_{1/b}^{p^+}\frac{d{\bar \mu}}{\bar \mu}
\lambda_{K}(\alpha_s({\bar \mu}))
+\frac{\alpha_s(p^+)}{\pi}{\cal C}_F\ln \nu\right)\right]
\nonumber \\
& &\times \phi(\xi,b,\mu)\;,
\label{sph}
\end{eqnarray}
We have set the lower limit of the variable $p^+$ to $1/b$, and the upper 
limit to $Q$, which is of the same order as $p^+_{\rm max}=Q/\sqrt{2}$. The 
initial condition $\phi(\xi,b,\mu)$ contains single ultraviolet logarithms 
from self-energy corrections to the valence quarks, which can be further 
organized through the RG equation,
\begin{equation}
{\cal D}\phi(\xi,b,\mu)=-2\lambda_q \phi(\xi,b,\mu)\;,
\label{uph}
\end{equation}
with 
\begin{equation}
{\cal D}=\mu\frac{\partial}{\partial \mu}+\beta(g)\frac{\partial}
{\partial g}
\end{equation}
and $\lambda_q=-\alpha_s/\pi$ the quark anomalous dimension 
in axial gauge. The solution to Eq.~(\ref{uph}) is given by
\begin{equation}
\phi(\xi,b,\mu)=
\exp\left[-2\int_{1/b}^{\mu}\frac{d {\bar \mu}}
{{\bar \mu}}\lambda_q (\alpha_s({\bar \mu})) \right]\phi(\xi,b,1/b)\;.
\label{uphs}
\end{equation}

Combining Eqs.~(\ref{sph}) and (\ref{uphs}), and employing the variable
changes $p^+=yQ$ in the first integral and ${\bar \mu}^2=yQ^2$ in the 
second integral, we resum the large radiative corrections
in the distribution function $\phi$ up to next-to-leading logarithms:
\begin{eqnarray}
\phi(\xi,p,b,\mu)&=&
\exp\left[-\int_{1/(bQ)}^1\frac{d y}{y}
\left(\int_{1/b^2}^{y^2Q^2}\frac{d{\bar \mu}^2}{{\bar \mu}^2}
\lambda_{K}(\alpha_s({\bar \mu}))
+2\frac{\alpha_s(yQ)}{\pi}{\cal C}_F\ln \nu\right)\right.
\nonumber \\
& &\left.-\int_{1/(bQ)^2}^{\mu^2/Q^2}\frac{d y}
{y}\lambda_q (\alpha_s(\sqrt{y}Q)) \right]
\phi(\xi,b,1/b)\;.
\label{coph}
\end{eqnarray}
\vskip 0.5cm

\noindent
III. 3 The jet function $J$

We now discuss the resummation in the case of the jet function $J$. There 
exists a similar chain rule for $J$,
\begin{equation}
p'^-\frac{d}{dp'^-}J=-\frac{n^2}{v'\cdot n}v'_\alpha\frac{d}
{dn_\alpha}J\;.
\label{cj}
\end{equation}
Employing the same procedures as in Sec. II, we obtain the derivative
$p'^-dJ/dp'^-$ described by Fig.~10(a), where the coefficient 2, again,
comes from the two outer most ends of the outgoing quark line, and the 
square represents the new vertex
\begin{equation}
{\hat v}'_\alpha=\frac{n^2v'_\alpha}{v'\cdot nn\cdot l}\;.
\label{nvp}
\end{equation}
The subdiagram containing the new vertex is factorized in the leading
soft and ultraviolet regions into a function $K'$ and a function $G'$,
respectively, whose general diagrams of $K'$ and $G'$ are shown in 
Fig.~10(b).

The resummation of large radiative corrections in $J$ is subtler. We argue 
that the resummation for $\phi$ results in Sudakov suppression in the large 
$b$ region, and thus $b$ can be set to zero approximately in the analysis of 
$J$. Another reason for this approximation is that the function $K'$ can 
be made free of infrared poles without introducing a $b$. This is obvious 
from the $O(\alpha_s)$ expression of $K'$ from Fig.~10(c),
\begin{eqnarray}
\delta(p'^2)K'&=&ig^2{\cal C}_F\mu^\epsilon\int
\frac{d^{4-\epsilon}l}{(2\pi)^{4-\epsilon}}
\frac{{\hat v}'_\mu v'_\nu}{v'\cdot l}
\left[\frac{\delta(p'^2)}{l^2}+2\pi i\delta(l^2)\delta((p'-l)^2)\right]
N^{\mu\nu}
\nonumber \\
& &-\delta(p'^2)\delta K'\;,
\label{kj}
\end{eqnarray}
where the $\delta$ functions correspond to the final-state cut, and
$\delta K'$ is an additive counterterm. It is easy to observe that $p'^-$ 
in the second term regulates the infrared divergence of the first integral. 
In order to extract the on-shell pole of the outgoing quark, we apply the 
operator
\begin{equation}
\delta(p'^2)\int d p'^2=\delta(p'^2)(2p'^-)
\int_0 ^{p'^+_{\rm max}} d p'^+
\label{pole}
\end{equation}
to the second term of the loop integral. In the case of $K'$ we choose 
$p'^+_{\rm max}=p'^-$. After a tedious calculation as exhibited in 
the Appendix, we have
\begin{equation}
K'=-\frac{\alpha_s}{\pi}{\cal C}_F\ln\frac{\nu'\mu}{p'^-}
\label{okj}
\end{equation}
with $\nu'=\sqrt{(v'\cdot n)^2/|n^2|}$ the gauge factor. 

The $O(\alpha_s)$ expression of $G'$ from Fig.~10(d) is given by
\begin{eqnarray}
G'&=&-ig^2{\cal C}_F\mu^\epsilon\int\frac{d^{4-\epsilon}l}
{(2\pi)^{4-\epsilon}}{\hat v}'_\mu
\left[\gamma_\nu\frac{\not p'-\not l}{(p'-l)^2}
+\frac{v'_\nu}{v'\cdot l}\right]\frac{N^{\mu\nu}}{l^2}
\nonumber \\
& &-\frac{\alpha_s}{\pi}{\cal C}_F\ln\left(\frac{1}{\nu'}
\frac{p^+}{\sqrt{p^+p'^-}}\right)-\delta G'\;,
\label{gj}
\end{eqnarray}
where $\delta G'$ is an additive counterterm. In the above formula the loop 
integral comes from the first two diagrams in Fig.~10(d), and the second 
term from the third and fourth diagrams. We obtain the second term by 
simply choosing $p'^+_{\rm max}=p^+\sim Q$ in Eq.~(\ref{pole}), which is 
greater than $p'^+_{\rm max}=p'^-$ for the soft function $K'$, and by 
choosing $\sqrt{q^2}\sim\sqrt{p^+p'^-}$, which is the typical scale of the 
third diagram, as the infrared cutoff of the integral for the fourth 
diagram. 
A similar calculation leads to
\begin{equation}
G'=-\frac{\alpha_s}{\pi}{\cal C}_F\ln\frac{\sqrt{p^+p'^-}}{\mu}\;.
\end{equation}

Hence, the general differential equation for $J$ is written as
\begin{eqnarray}
p'^-\frac{d}{dp'^-}J&=&2\left[K'(\nu'\mu/ p'^-,\alpha_s(\mu))+
G'(\sqrt{p^+p'^-}/\mu,\alpha_s(\mu))\right]J\;.
\label{dej}
\end{eqnarray}
The functions $K'$ and $G'$ for $J$ possess the same functional 
forms as $K$ and $G$ for $\phi$, respectively, and the counterterms 
have the relations $\delta K=\delta K'$ and $\delta G=\delta G'$.
We organize the single logarithms in $K'$ and $G'$ to give
\begin{eqnarray}
& &K'(\nu'\mu/p'^-,\alpha_s(\mu))+G(\sqrt{p^+p'^-}/\mu,\alpha_s(\mu))
\nonumber \\
& &\hskip 1.0cm
=-\frac{\alpha_s(p'^-)}{\pi}{\cal C}_F\ln \nu'-
\int_{p'^-}^{\sqrt{p^+p'^-}}\frac{d{\bar\mu}}{\bar\mu}
\lambda_K(\alpha_s({\bar\mu}))\;.
\label{skgj}
\end{eqnarray}
Solving Eq.~(\ref{dej}), we obtain 
\begin{eqnarray}
J(p',b,\mu)&=&
\exp\left[-2\int_{1/b}^{Q}\frac{d p'^-}{p'^-}
\left(\int_{p'^-}^{\sqrt{p^+p'^-}}\frac{d{\bar \mu}}{\bar \mu}
\lambda_{K}(\alpha_s({\bar \mu}))\right.\right.
\nonumber\\
& &\left.\left.+\frac{\alpha_s(p'^-)}{\pi}{\cal C}_F\ln \nu'\right)\right]
J(b,\mu)\;.
\label{coj1}
\end{eqnarray}

The initial condition $J(b,\mu)$ obeys a RG equation similar to
Eq.~(\ref{uph}), and is thus given by 
\begin{equation}
J(b,\mu)=
\exp\left[-\int_{Q/b}^{\mu^2}\frac{d {\bar \mu}^2}{{\bar \mu}^2}
\lambda_q (\alpha_s({\bar \mu})) \right]J(b,\sqrt{Q/b})\;,
\label{ujs}
\end{equation}
which sums the single ultraviolet logarithms arising from self-energy 
corrections to the outgoing quark. We have chosen the invariant mass of 
the jet $p'^2\sim p'^-p'^+\sim Q/b$ as the lower bound of ${\bar \mu}^2$ 
in the above integral. Combining Eqs.~(\ref{coj1}) and (\ref{ujs}), we 
resum the large radiative corrections in the jet function $J$ up to 
next-to-leading logarithms:
\begin{eqnarray}
J(p',b,\mu)&=&
\exp\left[-\int_{1/(bQ)}^1\frac{d y}{y}
\left(\int_{y^2Q^2}^{yQ^2}\frac{d{\bar \mu}^2}{{\bar \mu}^2}
\lambda_{\cal K}(\alpha_s({\bar \mu}))
+2\frac{\alpha_s(yQ)}{\pi}{\cal C}_F\ln \nu'\right)\right.
\nonumber \\
& &\left.-\int_{1/(bQ)}^{\mu^2/Q^2}\frac{d y}
{y}\lambda_q (\alpha_s(\sqrt{y}Q)) \right]J(b,\sqrt{Q/b})\;.
\label{coj}
\end{eqnarray}
where the variable changes similar to those for $\phi$ have been employed.
Note that we have taken into account four scales in the resummation 
procedures for $J$, which have the ordering 
$Q>\sqrt{p^+p'^-}>p'^->1/b>\Lambda_{\rm QCD}$. It is easy to observe that
the integral generating double logarithms vanishes in the 
$y\to 1$, {\it ie.}, $p'\gg 1/b$, region, where the jet is energetic.
\vskip 0.5cm

\noindent
III. 4 The soft function $U$ and the hard scattering $H$

We have organized the single ultraviolet logarithms in the distribution
function $\phi$ and in the jet function $J$ into the exponentials in
terms of the anomalous dimension $\lambda_q$. To sum up next-to-leading 
logarithms, we must include the single soft logarithms in $U$. We shall
show that the gauge factors $\nu$ and $\nu'$ in $\phi$ and 
$J$, respectively, are canceled by those in $U$. 

The RG equation for $U$ is written as
\begin{equation}
{\cal D}U(b,\mu)=-\lambda_U U(b,\mu)\;,
\label{us}
\end{equation}
where the anomalous dimension $\lambda_U$ is derived from the first 
diagram of Fig.~7(c). The one-loop bare soft function $U^{(0)}$ is given by
\begin{eqnarray}
U^{(0)}&=&-2ig^2{\cal C}_F\mu^\epsilon\int\frac{d^{4-\epsilon}l}
{(2\pi)^{4-\epsilon}}\frac{v_\mu v'_\nu N^{\mu\nu}}
{v\cdot lv'\cdot l l^2}\;,
\nonumber \\
&=&-2\frac{\alpha_s}{\pi}\ln(\nu\nu')\frac{1}{\epsilon}\;.
\label{udi}
\end{eqnarray}
The coefficient 2 in the first expression corresponds to 
the two vertex-correction diagrams, and those constants of order unity 
irrelevant to the gauge dependence have been neglected in the second 
expression. The gauge dependent factor $\ln \nu$ ($\ln \nu'$) in 
Eq.~(\ref{udi}) comes from the term $n^\mu l^\nu/n\cdot l$ 
($n^\nu l^\mu/n\cdot l$) in $N^{\mu\nu}$.

It is then trivial to compute $\lambda_U$ from the multiplicative 
renormalization constant $Z_U$, defined by $U^{(0)}=Z_U U$, which is 
written as
\begin{equation}
\lambda_U=-2\frac{\alpha_s}{\pi}{\cal C}_F\ln(\nu\nu')\;.
\label{au}
\end{equation}
We obtain the solution to Eq.~(\ref{us}) 
\begin{equation}
U(b,\mu)=\exp\left[-\int_{1/(bQ)}^{\mu/Q}\frac{dy}
{y}\lambda_U (\alpha_s(yQ)) \right]U(b,1/b)\;.
\label{uu}
\end{equation}

Since the structure function $F$ is independent of $\mu$, we have the RG 
equation for the hard scattering $H$ 
\begin{equation}
{\cal D}H(\xi,Q,\mu)=(4\lambda_q+\lambda_U)H(\xi,Q,\mu)\;,
\label{sh}
\end{equation}
which is solved to give
\begin{eqnarray}
H(\xi,Q,\mu)&=&\exp\left[-2\int_{\mu^2/Q^2}^{1}\frac{dy}{y}
\lambda_q(\alpha_s(\sqrt{y}Q)) -\int_{\mu/Q}^{1}\frac{dy}{y}
\lambda_U (\alpha_s(yQ)) \right]
\nonumber \\
& &\times H(\xi,Q,Q)\;.
\label{uh}
\end{eqnarray}
In the above expressions for $U$ and $H$ necessary variable changes
have been made.

Substituting Eqs.~(\ref{coph}), (\ref{coj}), (\ref{uu}) and (\ref{uh})
into (\ref{fdis}), we derive
\begin{eqnarray}
F(x,Q^2)&=&\int_x^1 d\xi \int d^2{\bf b}\phi(\xi,b,1/b)J(b,\sqrt{Q/b})
U(b,1/b)H(\xi,Q,Q)
\nonumber \\
& &\times \exp[-S_{\rm DIS}(Q,b)]\;,
\label{fdiss}
\end{eqnarray}
where the Sudakov exponent $S_{\rm DIS}$ is given by
\begin{eqnarray}
S_{\rm DIS}&=&\int_{1/(bQ)}^1\frac{d y}{y}\left[
\int_{1/b^2}^{yQ^2}\frac{d{\bar \mu}^2}{{\bar \mu}^2}
\lambda_{K}(\alpha_s({\bar \mu}))+\lambda_q(\alpha_s(\sqrt{y}Q))\right]
\nonumber\\
& &+\int_{1/(bQ)^2}^1\frac{d y}{y}\lambda_q(\alpha_s(\sqrt{y}Q))\;.
\label{fs1}
\end{eqnarray}
It is obvious that the gauge factors $\nu$ from $G$ and $\nu'$
from $K'$ have been canceled by those from $U$, and that the resummation 
technique developed above is gauge independent. This is an issue which  
cannot be addressed in the conventional Wilson-loop approach, since the 
vector $n$ has been fixed in the direction $v$ or $v'$ in order to form
a loop along the classical trajectories of the quarks. It is easy to 
observe that Eq.~(\ref{fs1}) is the same as the corresponding 
formulas derived from the Wilson-loop formalism with the identification 
of the anomalous dimension $\lambda_K=A$ in \cite{CT} and $\lambda_K=
\Gamma_{\rm cusp}$ in \cite{KM}. Note that the second integral involving
$\lambda_q$ in Eq.~(\ref{fs1}) does not exist in \cite{CT,KM}, because 
the self-energy corrections to the valence quarks can not be taken into 
account by a Wilson (eikonal) line. The self-energy corrections to the 
outgoing quark, denoted by the second term in the first integral, also 
appear in \cite{CT,KM}, which were introduced, however, through an 
Altarelli-Parisi evolution equation.

\vskip 2.0cm

\centerline{\large \bf IV. Drell-Yan Production}
\vskip 0.5cm

In this section we discuss the resummation for Drell-Yan production, 
\begin{equation}
h(p_1)+h(p_2) \to \ell{\bar \ell}(q) + X\;,
\label{dy}
\end{equation}
where $p_1=(p_1^+,0,{\bf 0}_T)$ and $p_2=(0,p_2^-,{\bf 0}_T)$ with $p_1^+= 
p_2^-$ are the momenta of the incoming hadrons $h$ in their center-of-mass 
frame, and $q$ is the momentum of the lepton pair produced in the  
process.

We start with the annihilation at the quark level as shown in Fig.~11(a),
and then consider $O(\alpha_s)$ corrections from Figs.~11(b)-(e). 
Figs.~11(b) and 11(c) have been analyzed in Sec. III, and the conclusion 
is the same: They produce double logarithms with
collinear ones dominant, and are thus absorbed into distribution functions
$\phi_1$ or $\phi_2$. The treatment of Figs.~11(d) and 11(e) is
similar to that of Fig.~6(e). If the loop momentum $l$ is parallel to
$p_1$, we replace the second quark by an eikonal line in the direction
$n$, and factorize the gluon into $\phi_1$. Certainly, a soft subtraction
with the first quark further replaced by an eikonal line in the direction 
$v_1=p_1/p_1^+$, similar to Fig.~7(a), is necessary. If $l$ is parallel to 
$p_2$, the diagram is factorized into $\phi_2$ following the above
prescription. If $l$ is soft, the first and second quarks are replaced by 
eikonal lines along $v_1$ and $v_2=p_2/p_2^-$, respectively, 
and the diagram is factorized into a soft function $U$. A corresponding 
collinear subtraction similar to Fig.~7(c), should be included. 

In axial gauge $n\cdot A=0$ contributions from all the above diagrams 
vanish, except for the one with both quarks replaced by eikonal lines 
along their classical trajectories. At last, when $l$ is hard, 
Figs.~11(b)-(e) are absorbed into a hard scattering amplitude 
$H$. Hence, we arrive at the general factorization picture for the cross 
section of a Drell-Yan process as shown in Fig.~12, and the corresponding 
formula,
\begin{eqnarray}
Q^2\frac{d\sigma(\tau,Q^2)}{dQ^2}&=&
\int d\xi_1 d\xi_2 \int d^2{\bf b}
\phi_1(\xi_1,p_1,b,\mu)\phi_2(\xi_2,p_2,b,\mu)
\nonumber \\
& &\times U(b,\mu)H(\xi_1,\xi_2,Q,\mu)\;,
\label{fdy}
\end{eqnarray}
where the variable $\tau$ is defined by \cite{CT}
\begin{equation}
\tau=\frac{Q^2}{s}\;,
\end{equation}
with $s=(p_1+p_2)^2$ and $Q^2=q^2$ the invariant mass of the lepton pair. 
We have assumed that the first (second) quark carries a fractional momentum 
$\xi_1p_1^++{\bf p}_{1T}$ ($\xi_2p_2^-+{\bf p}_{2T}$) with a small 
transverse momentum $p_{1T}$ ($p_{2T}$), which regulates the infrared 
divergences of higher-order corrections. The cross section $\sigma$ 
depends only on a single $b=b_1=b_2$, where $b_1$ and $b_2$ are the 
conjugate variables of $p_{1T}$ and $p_{2T}$, respectively. The $b$ 
dependence comes from the Fourier factor $e^{i{\bf l}_T\cdot {\bf b}}$, 
which is associated with a loop integrand whenever the loop momentum 
flows through $\phi_1$ or $\phi_2$.

We discuss the resummation of large radiative corrections in Drell-Yan  
production. We emphasize that double logarithms exist for arbitrary
intermediate and large values of $\tau$, and no new double logarithms  
occur in the end-point region with $\tau\to 1$. This differs from the 
case of deeply inelastic scattering, where extra double logarithms appear
at the end point. In order to compare the results for 
deeply inelastic scattering and for Drell-Yan production, we concentrate 
on the $\tau\to 1$ region, where both $\xi_1$ and $\xi_2$ approach unity.
The resummation for the distribution functions $\phi_1$ and $\phi_2$ is 
the same as that in deeply inelastic scattering. We quote Eq.~(\ref{coph}) 
directly here:
\begin{eqnarray}
\phi_i(\xi_i,p_i,b,\mu)&=&
\exp\left[-\int_{1/(bQ)}^1\frac{d y}{y}
\left(\int_{1/b^2}^{y^2Q^2}\frac{d{\bar \mu}^2}{{\bar \mu}^2}
\lambda_{K}(\alpha_s({\bar \mu}))
+2\frac{\alpha_s(yQ)}{\pi}{\cal C}_F\ln \nu_i\right)\right.
\nonumber \\
& &\left.-\int_{1/(bQ)^2}^{\mu^2/Q^2}\frac{d y}
{y}\lambda_q (\alpha_s(\sqrt{y}Q)) \right]
\phi_i(\xi_i,b,1/b)\;,
\label{cophd}
\end{eqnarray}
for $i=1$, 2, where the gauge factors $\nu_i$ are defined
by $\nu_i=\sqrt{(v_i\cdot n)^2/|n^2|}$.

We then consider the soft function $U$. The lowest-order diagrams
contributing to $U$ are those with a gluon attaching two eikonal lines on 
the same side of the final-state cut. The diagrams with a real gluon 
attaching two eikonal lines on the different sides of the final-state cut 
are ultravioletly finite due to the factor $e^{i{\bf l}_T\cdot {\bf b}}$
associated with the corresponding loop integrand. Hence, the anomalous 
dimension
\begin{equation}
\lambda_U=-2\frac{\alpha_s}{\pi}{\cal C}_F\ln(\nu_1\nu_2)
\label{aud}
\end{equation}
is similar to that for deeply inelastic scattering in Eq.~(\ref{au}). The 
evolution of the soft function is then described by Eq.~(\ref{uu}). 
Accordingly, the summation of the single ultraviolet logarithms in $H$ 
leads to an evolution similar to Eq.~(\ref{uh}).

Combining the above results for each convolution factor, Eq.~(\ref{fdy}) 
is reexpressed as
\begin{eqnarray}
Q^2\frac{d\sigma(\tau,Q^2)}{dQ^2}&=&
\int d\xi_1 d\xi_2 \int d^2{\bf b}
\phi_1(\xi_1,b,1/b)\phi_2(\xi_2,b,1/b)
\nonumber \\
& &\times U(b,1/b)H(\xi_1,\xi_2,Q,Q)\exp[-S_{\rm DY}(Q,b)]\;,
\label{fdy1}
\end{eqnarray}
where the Sudakov exponent $S_{\rm DY}$ is given by
\begin{equation}
S_{\rm DY}=2\int_{1/(bQ)}^1\frac{d y}{y}
\int_{1/b^2}^{y^2Q^2}\frac{d{\bar \mu}^2}{{\bar \mu}^2}
\lambda_{K}(\alpha_s({\bar \mu}))
+2\int_{1/(bQ)^2}^1\frac{d y}{y}
\lambda_q(\alpha_s(\sqrt{y}Q))\;.
\label{fs2}
\end{equation}
Again, the gauge factors $\nu_i$ from $G$ have been canceled by those from 
$U$, such that the resummation technique is gauge independent. 

We consider the ratio $\Delta=e^{-S_{\rm DY}}/(2e^{-S_{\rm DIS}})$,
\begin{eqnarray}
\Delta=\exp\left[2\int_{1/(bQ)}^1\frac{d y}{y}
\left(\int_{y^2Q^2}^{yQ^2}\frac{d{\bar \mu}^2}{{\bar \mu}^2}
\lambda_{K}(\alpha_s({\bar \mu}))
-\frac{3}{2}\frac{{\cal C}_F}{\pi}\alpha_s(\sqrt{y}Q)\right) \right]\;,
\label{de}
\end{eqnarray}
which is similar to the coefficient function $\Delta_N$ in \cite{CT},
\begin{eqnarray}
\Delta_N&=&\exp\left[2\int_0^1dx\frac{1-x^{N-1}}{1-x}
\left(\int_{(1-x)^2Q^2}^{(1-x)Q^2}\frac{d{\bar \mu}^2}{{\bar \mu}^2}
\lambda_{K}(\alpha_s({\bar \mu}))\right.\right.
\nonumber \\
& &\left.\left.-\frac{3}{2}\frac{{\cal C}_F}{\pi}
\alpha_s(\sqrt{1-x}Q)\right) \right]\;,
\label{den}
\end{eqnarray}
with $bQ$ corresponding to the large number $N$. $\Delta_N$ is associated
with the evolution of the $N$-th moment of the differential cross section.
Eq.~(\ref{de}) is exactly the resummation result of the jet function $J$ 
in deeply inelastic scattering, excluding the gauge dependent factor. 
We conclude that the resummation technique is equivalent to the
Wilson-loop approach in the study of deeply inelastic scattering and 
Drell-Yan production.

\vskip 2.0cm

\centerline{\large \bf V. INCLUSIVE HEAVY MESON DECAYS}
\vskip 0.5cm

The technique developed above can be applied to inclusive heavy meson 
decays, such as $B\to X_s\gamma$ and $B\to X_u l\nu$, straightforwardly. 
The resummation of large radiative corrections in these processes at the 
end points of spectra \cite{BSUV,ACC,FJM} has been performed in 
\cite{L1,LY1}, where the specific gauge vectors $n\propto (1,1,{\bf 0}_T)$ 
and $n\propto (1,-1,{\bf 0}_T)$ were employed, respectively. In this 
section we shall discuss the case with a general $n$, and show that the \
results are gauge independent.

The Wilson-loop formalism has been applied to the study of end-point 
singularities in inclusive $B$ meson decays \cite{KS} based on the relation
between Wilson lines and the heavy quark effective theory (HQET) \cite{G} 
explored in \cite{KR2}. The Wilson loop is absorbed into a heavy quark 
distribution function, whose evolution is governed by RG equations. We 
shall demonstrate that the resummation technique is equivalent to the 
Wilson-loop formalism in the study of inclusive heavy meson decays. 

We concentrate on the inclusive radiative decay $B\to X_s\gamma$, which 
occurs through the transition $b\to s\gamma$ described by an effective 
Hamiltonian \cite{SVZ}. The basic factorization of this process at the 
quark level is similar to Fig.~6(a). The momentum of the $b$ quark is $P-k$, 
$P=(M/\sqrt{2})(1,1,{\bf 0}_T)$ being the $B$ meson momentum, and $k$ being 
the momentum carried by light partons in the $B$ meson. $k$ has a plus 
component $k^+$ and small transverse components ${\bf k}_T$. These momenta 
satisfy the on-shell conditions $(P-k)^2\approx M_b^2$ and $k^2\approx 0$. 
Here $M$ and $M_b$ are the $B$ meson mass and the $b$ quark mass, 
respectively. The $b$ quark decays into a real photon of momentum $q$ and a 
$s$ quark of momentum $p'$, which is regarded as being light. Assuming that 
the nonvanishing component of $q$ is $q^+$, we have $p'=P-k-q=
(P^+-k^+-q^+,P^-,-{\bf k}_T)$.

The absorption of infrared divergences is analyzed below. The self-energy 
correction to the $b$ quark and the loop correction with a real gluon 
connecting the two $b$ quarks as in Fig.~6(b) and 6(c), respectively,
which contain only single soft logarithms, are grouped into a distribution 
function $\phi$. In the end-point region with $q^+\to P^+$, the $s$ quark 
has a large minus component $p'^-=P^-$ but a small invariant mass $p'^2$. 
The case is then similar to deeply inelastic scattering at the end point, 
and thus extra double logarithms appear. Therefore, the self-energy 
correction to the $s$ quark in Fig.~6(d) is absorbed into a jet function 
$J$. 

The photon vertex correction gives both collinear logarithms 
from the loop momentum $l$ parallel to $p'$ in the end-point region and 
soft logarithms from vanishing $l$. If $l$ is parallel to $p'$, we replace
the $b$ quark by an eikonal line in the direction $n$, and factorize 
the diagram into $J$. In axial gauge $n\cdot A=0$ this diagram along 
with its soft subtraction do not contribute. If $l$ is soft, the 
$b$ and $s$ quarks are replaced by eikonal lines along $v=\sqrt{2}P/M$ 
and $v'=p'/p'^-$, respectively \cite{L1}. The replacement of 
a $b$ quark by an eikonal line is consistent with the flavor symmetry 
in the HQET. We then absorb the gluon 
into a soft function $U$. Similarly, the collinear subtraction 
associated with this diagram does not contribute in axial gauge.

Remaining important corrections from hard gluons are absorbed into a 
hard scattering amplitude $H$. At last, the factorization formula for 
the spectrum of the decay $B\to X_s\gamma$ in $b$ space is written as
\begin{equation}
\frac{d \Gamma_\gamma}{d E_\gamma}=\int d \xi \int d^2 {\bf b}
\phi(\xi,b,\mu)J(p',b,\mu)U(b,\mu)H(\xi,M,\mu)\;,
\label{facb}
\end{equation}
which is also described by Fig.~8. Here $b$ is the conjugate variable of
$k_T$, $E_\gamma=q^+/\sqrt{2}$ is the photon energy, $\xi=1-k^+/P^+$ is 
the momentum fraction, and $\mu$ is the factorization and renormalization 
scale. The resummation for $J$ has been performed in Sec. II, and the 
result is the same as Eq.~(\ref{coj}) but with $M$ substituted for $Q$.

The single soft logarithms in $\phi$ can be collected by eikonal lines in 
the direction $v$ \cite{LY1}. The self-energy correction to the $b$ quark 
vanishes under this eikonal approximation. Hence, we consider only the 
diagram with a real gluon attaching the two eikonal lines, which leads to 
the loop integral for the bare distribution function $\phi^{(0)}$,
\begin{equation}
\phi^{(0)}=g^2{\cal C}_F\mu^\epsilon\int\frac{d^{4-\epsilon}l}
{(2\pi)^{4-\epsilon}}\frac{v_\mu v_\nu N^{\mu\nu}}{(v\cdot l)^2}
2\pi\delta(l^2)e^{i{\bf l}_T\cdot {\bf b}}\;.
\label{p1}
\end{equation}
Evaluating Eq.~(\ref{p1}), we obtain the pole term, and thus the
anomalous dimension of $\phi$.

Similarly, the bare soft function $U^{(0)}$ is written as
\begin{equation}
U^{(0)}=-2ig^2{\cal C}_F\mu^\epsilon\int\frac{d^{4-\epsilon}l}
{(2\pi)^{4-\epsilon}}\frac{v_\mu v'_\nu N^{\mu\nu}}
{v\cdot l v'\cdot l l^2}\;,
\label{u1}
\end{equation}
where the two vertex-correction diagrams have been included. The term
$-n^\nu l^\mu/(n\cdot l)$ in $N^{\mu\nu}$ gives the gauge dependent 
factor $\ln\nu'$, which cancels that of $J$ as in the case of deeply
inelastic scattering. The contribution from the term $n^\mu l^\nu/
(n\cdot l)$ cancels those from $(n^\mu l^\nu+n^\nu l^\mu)/(n\cdot l)$ in 
$\phi^{(0)}$. The other two terms $g^{\mu\nu}$ and $n^2 l^\mu l^\nu/
(n\cdot l)^2$ lead to gauge-independent results of order unity, which are 
thus neglected.

Based on the above analysis, we need to compute only the contributions
to $\phi^{(0)}$ from $g^{\mu\nu}$ and from $n^2l^\mu l^\nu/(n\cdot l)^2$.
Details of the calculation refer to the Appendix. The anomalous dimension
\begin{equation}
\lambda_\phi=-\frac{\alpha_s}{\pi}{\cal C}_F
\label{php}
\end{equation}
is then obtained, which is independent of $n$. The evolution of $\phi$ is 
given by 
\begin{equation}
\phi(\xi,b,\mu)=\exp\left[-\int_{1/(bM)}^{\mu/M}\frac{d y}
{y}\lambda_\phi (\alpha_s(yM)) \right]\phi(\xi,b,1/b)\;.
\end{equation}

Combining the evolution of the hard scattering to the largest scale $M$, 
\begin{eqnarray}
H(\xi,M,\mu)&=&\exp\left[-\int_{\mu/M}^1\frac{d y}{y}
\lambda_\phi(\alpha_s(yM)) -\int_{\mu^2/M^2}^1\frac{d y}{y}
\lambda_q(\alpha_s(\sqrt{y}M)) \right]
\nonumber \\
& &\times H(\xi,M,M)\;,
\end{eqnarray}
we derive 
\begin{eqnarray}
\frac{d \Gamma_\gamma}{d E_\gamma}&=&\int d \xi \int d^2 {\bf b}
\phi(\xi,b,1/b)J(p',b,\sqrt{M/b})U(b,1/b)H(M,M)
\nonumber\\
& &\times \exp[-S_\gamma(M,b)]\;,
\label{facb1}
\end{eqnarray}
with the Sudakov exponent 
\begin{eqnarray}
S_\gamma=\int_{1/(bM)}^1\frac{d y}{y}
\left[\int_{y^2M^2}^{yM^2}\frac{d{\bar \mu}^2}{{\bar \mu}^2}
\lambda_{K}(\alpha_s({\bar \mu}))+\lambda_\phi (\alpha_s(yM)) 
+\lambda_q (\alpha_s(\sqrt{y}M)) \right]\;.
\label{bs}
\end{eqnarray}
Note that the above expression differs from that employed in the 
PQCD analysis of inclusive $B$ meson decays \cite{LY1}, where the lower
and upper bounds of the variable ${\bar \mu}$ are set to $1/b$ and $yM$,
respectively. The distinction is attributed to the different ways to
regulate the infrared divergences in $K'$: The function $\delta((p'-l)^2)$ 
was introduced in the present work, while the Fourier factor
$e^{i{\bf l}_T\cdot {\bf b}}$ was introduced in \cite{LY1}, such that
the typical scales of $K'$ are $p'^-$ and $1/b$, respectively.

For comparision, we exhibit the result from the Wilson-loop approach
\cite{KS},
\begin{equation}
S_N=\int_{n_0/N}^{1}\frac{d y}{y}\left[
\int_{y^2M^2}^{yM^2}\frac{d k_t^2}{k_t^2}
\Gamma_{\rm cusp}(\alpha_s(k_t))+\Gamma(\alpha_s(yM))
+\gamma(\alpha_s(\sqrt{y}M))\right]\;,
\label{fks}
\end{equation}
where $S_N$ describes the evolution of the $N$-th moment of the differential
decay rate, with the $b$ quark mass $M$, the constant $n_0=e^{-\gamma_E}$,
$\gamma_E$ being the Euler constant. The anomalous dimensions are given by
$\Gamma_{\rm cusp}=\lambda_{K}$, $\Gamma=-(\alpha_s/\pi){\cal C}_F$ and 
$\gamma=-\alpha_s/\pi$ \cite{KM,KR1}. From Eqs.~(\ref{bs}) and (\ref{fks}),
we easily identify the equality of the anomalous dimensions: 
\begin{eqnarray}
\lambda_{K}=\Gamma_{\rm cusp}\;,\;\;\;\;
\lambda_\phi=\Gamma\;,\;\;\;\;\lambda_q=\gamma\;.
\label{re}
\end{eqnarray}
The two exponents $S_\gamma$ and $S_N$ are basically the same except for 
the lower bounds of $y$. The scale $1/b$ comes from the inclusion of 
transverse degrees of freedom in our formalism, and the $N$ dependence is 
due to the moment analysis in \cite{KS}.
\vskip 2.0cm

\centerline{\large \bf VI. CONCLUSION}
\vskip 0.5cm

In this paper we have discussed the resummation of large radiative 
corrections in inclusive processes at the end points, such as deeply
inelastic scattering, Drell-Yan production and inclusive $B$ meson decays,
which lead to the same results as those from the conventional 
Wilson-loop formalism. We have demonstrated the equivalence of
the two very different approaches by carefully choosing the infrared cutoffs 
for the jet functions in deeply inelastic scattering and in inclusive 
radiative $B$ meson decays. Furthermore, we have shown in details 
that the resummation technique is gauge invariant: It can be performed both
in covariant gauge and in axial gauge with an arbitrary gauge vector. The
gauge dependent factors in the distribution functions and in the jet
functions are canceled by those in the soft functions.

We emphasize that a complete gauge-invariant set of higher-twist diagrams 
contains those with valence partons carrying transverse momenta as 
considered in this work, and those with more partons entering a hard 
scattering. We observe that the sole inclusion of the former in the study 
of inclusive processes at the end points is sufficient to guarantee the 
gauge invariance. Take the distribution function in deeply inelastic 
scattering as an example. The fractional parton momentum $\xi Q$, which is 
the upper bound of the variable $p^+$ in the resummation integral, 
approaches the largest scale $Q$ at the end point. The upper bound of the 
scale $\mu$ in the evolution integral for the hard scattering is also $Q$, 
such that the cancellation of the gauge dependent factors is exact. 

In the region away from the end points there then exist residual gauge 
dependent factors in the form
\begin{equation}
\exp\left(\frac{{\cal C}_F}{\pi}\ln \nu\int_{\xi Q}^Q\frac{d \mu}{\mu}
\alpha_s(\mu)\right)\;.
\label{gau}
\end{equation}
In order to maintain the gauge invariance, at least the gauge-dependent 
contributions from multiparton diagrams should be able to be resummed into 
an exponential, such that Eq.~(\ref{gau}) can be canceled. This 
observation hints that it is possible to develop a new resummation 
technique for higher-twist effects from multiparton diagrams. We shall 
discuss this issue in a separate work.

An immediate extension of the formalism presented here is to organize
large corrections in processes involving gluons that enter a hard 
scattering as partons. This extension will be very useful for the study
of, say, direct-photon processes, photoproduction and gluon fusion.
We shall investigate this subject elsewhere. 
\vskip 2.0cm

\centerline{\large\bf Acknowledgement}
\vskip 0.5cm

This work was supported by the National Science Council of R.O.C. under 
Grant No. NSC85-2112-M194-009.

\newpage

\centerline{\large {\bf APPENDIX}}
\vskip 0.5cm
In this appendix we present some details of the one-loop evaluation
of the functions $K$, $G$, $K'$, $G'$ and the anomalous dimension 
$\lambda_\phi$.

\noindent
1. The function $K$ in Eq.~(\ref{kph})

The function $K$ from the two diagrams in Fig.~9(c) is written as
\begin{eqnarray*}
{K}=ig^2{\cal C}_F\mu^\epsilon\int\frac{d^{4-\epsilon}l}
{(2\pi)^{4-\epsilon}}\left[\frac{1}{l^2}+2\pi i\delta(l^2)
e^{i{\bf l}_T\cdot {\bf b}}\right]\frac{n^2v_\mu v_\nu N^{\mu\nu}}
{v\cdot nn\cdot lv\cdot l}-\delta{K}\;.
\label{k}
\end{eqnarray*}
Consider the former part of the integral without the exponential factor,
\begin{eqnarray*}
{K}_1=ig^2{\cal C}_F\mu^\epsilon\int\frac{d^{4-\epsilon}l}
{(2\pi)^{4-\epsilon}}\frac{n^2v_\mu v_\nu N^{\mu\nu}}
{v\cdot nn\cdot lv\cdot l l^2}\;.
\end{eqnarray*}
The first term $g^{\mu\nu}$ in $N^{\mu\nu}$ does not contribute,
since it leads to a numerator $v^2=0$. The second and third terms,
$-n^\mu l^\nu/(n\cdot l)$ and $-n^\nu l^\mu/(n\cdot l)$, give
equal contributions, which are given by
\begin{eqnarray*}
{K}_{12}={K}_{13}
=-ig^2{\cal C}_F\mu^\epsilon n^2\int\frac{d^{4-\epsilon}l}
{(2\pi)^{4-\epsilon}}\frac{1}{l^2(n\cdot l)^2}\;.
\label{33}
\end{eqnarray*}
We evaluate the above formula as a contour integral, assuming $n^+<0$, 
$n^->0$ and ${\bf n}_T=0$ for simplicity. Certainly, the final expression 
should be Lorentz invariant. The reason for choosing a space-like $n$ will 
become clear in the evaluation of $G$. Performing the integration over
$l^+$, whose singularity is pinched for $l^-<0$, we obtain
\begin{eqnarray*}
{K}_{12}=g^2{\cal C}_F\mu^\epsilon n^2\int\frac{d^{2-\epsilon}l}
{(2\pi)^{3-\epsilon}}\int_{-\infty}^0d l^-\frac{2l^-}
{(2n^+l^{-2}+n^-l_T^2)^2}\;.
\end{eqnarray*}
Doing the $l^-$ integration straightforwardly, we arrive at
\begin{eqnarray*}
{K}_{12}=-g^2{\cal C}_F\mu^\epsilon \int\frac{d^{2-\epsilon}l}
{(2\pi)^{3-\epsilon}}\frac{1}{l_T^2+a^2}\;,
\label{k12}
\end{eqnarray*}
in which the small constant $a$ serves as an infrared cutoff, and will be
set to zero at last. Computing the above integral using the standard 
dimensional regularization, we derive
\begin{eqnarray*}
{K}_{12}=-\frac{\alpha_s}{2\pi}{\cal C}_F\left(\frac{2}{\epsilon}+
\ln\frac{4\pi\mu^2}{a^2e^{\gamma_E}}\right)\;.
\label{kp1}
\end{eqnarray*}

The fourth term in $N^{\mu\nu}$, $n^2l^\mu l^\nu/(n\cdot l)^2$,
leads to the integral
\begin{eqnarray*}
{K}_{14}&=&ig^2{\cal C}_F\mu^\epsilon \frac{(n^2)^2}{v\cdot n}
\int\frac{d^{4-\epsilon}l}{(2\pi)^{4-\epsilon}}\frac{v\cdot l}
{l^2(n\cdot l)^3}
\nonumber \\
&=&ig^2{\cal C}_F\mu^\epsilon \frac{(n^2)^2}{v\cdot n}
\left(-\frac{1}{2}\frac{d}{d n^-}\right)\int\frac{d^{4-\epsilon}l}
{(2\pi)^{4-\epsilon}}\frac{1}{l^2(n\cdot l)^2}\;,
\nonumber 
\end{eqnarray*}
where the integral to be evaluated is the same as $K_{12}$. It is then 
trivial to show ${K}_{14}=-{K}_{12}$, and thus ${K}_{1}={K}_{12}+{K}_{13}
+{K}_{14}={K}_{12}$. The second expression in Eq.~(\ref{k2}) also follows
this conclusion.

We now consider the latter part of the integral with the exponential 
factor,
\begin{eqnarray*}
{K}_2=-g^2{\cal C}_F\mu^\epsilon\int\frac{d^{4-\epsilon}l}
{(2\pi)^{3-\epsilon}}\delta(l^2)e^{i{\bf l}_T\cdot {\bf b}}
\frac{n^2v_\mu v_\nu N^{\mu\nu}}{v\cdot nn\cdot lv\cdot l}\;.
\end{eqnarray*}
Following the above procedures, we find that only the contribution from 
the second term of $N^{\mu\nu}$ needs to be computed, which gives a 
similar expression to $K_{12}$,
\begin{eqnarray*}
{K}_{2}=g^2{\cal C}_F\mu^\epsilon \int\frac{d^{2-\epsilon}l}
{(2\pi)^{3-\epsilon}}\frac{1}{l_T^2+a^2}e^{i{\bf l}_T\cdot {\bf b}}\;.
\label{k22}
\end{eqnarray*}
Since this integral is free of ultraviolet and infrared singularities,
it can be evaluated by setting $\epsilon=0$ directly. We obtain
\begin{eqnarray*}
{K}_2=\frac{\alpha_s}{\pi}{\cal C}_FK_0(ab)
\approx\frac{\alpha_s}{2\pi}{\cal C}_F\ln\frac{4e^{-2\gamma_E}}{a^2b^2}
\end{eqnarray*}
in the limit $a\to 0$, where $K_0$ is the modified Bessel function
of zeroth order. The sum $K_1+K_2$ is then given by
\begin{eqnarray*}
{K}_1+{K}_2=-\frac{\alpha_s}{2\pi}{\cal C}_F\left[
\frac{2}{\epsilon}+\ln(\pi\mu^2b^2e^{\gamma_E})\right]\;,
\end{eqnarray*}
which is independent of the cutoff $a$. If renormalizing $K$ in 
${\overline {\rm MS}}$ scheme, {\it ie.}, 
choosing
\begin{eqnarray*}
\delta{K}=-\frac{\alpha_s}{2\pi}{\cal C}_F\left(\frac{2}{\epsilon}
+\ln 4\pi-\gamma_E\right)\;,
\label{dkp1}
\end{eqnarray*}
we have the lowest-order ${K}={K}_1+{K}_2-\delta{K}$ as
\begin{eqnarray*}
{K}=-\frac{\alpha_s}{\pi}{\cal C}_F\left[
\ln(b\mu)+\ln\frac{e^{\gamma_E}}{2}\right]\;.
\label{40}
\end{eqnarray*}
Neglecting the second term of order unity, we derive Eq.~(\ref{ok}).

\vskip 1.0cm

\noindent
2. The function $G$ in Eq.~(\ref{gph})

The function $G$ from Fig.~9(d) is written as
\begin{eqnarray*}
{G}=ig^2{\cal C}_F\mu^\epsilon\int\frac{d^{4-\epsilon}l}
{(2\pi)^{4-\epsilon}}\frac{n^2v_\mu N^{\mu\nu}}{v\cdot nn\cdot l l^2}
\left[\frac{\not p+\not l}{(p+l)^2}\gamma_\nu-\frac{v_\nu}{v\cdot l}
\right]-\delta{G}\;.
\nonumber
\end{eqnarray*} 
A simple investigation 
shows that only the third term $-n^\nu l^\mu/(n\cdot l)$ contributes. 
Hence, we express the integral as ${G}_{13}-{K}_{13}$, where 
${K}_{13}=K_{12}$ has been computed before, and ${G}_{13}$ is the 
former part of the integral,
\begin{eqnarray*}
{G}_{13}&=&-ig^2{\cal C}_F\mu^\epsilon\frac{n^2}{v\cdot n}
\int\frac{d^{4-\epsilon}l}{(2\pi)^{4-\epsilon}}
\frac{(\not p+\not l)v\cdot l}{(p+l)^2(n\cdot l)^2l^2}\not n
\nonumber \\
&=&-2ig^2{\cal C}_F\mu^\epsilon n^2
\int\frac{d^{4-\epsilon}l}{(2\pi)^{4-\epsilon}}
\frac{(p+l)^+l^-}{(p+l)^2(n\cdot l)^2l^2}\;.
\nonumber
\end{eqnarray*}
In the second expression those terms, vanishing after contracted with
the matrix structure of the initial hadron, have been neglected.
Performing the integration over $l^-$ and then over $l^+$, we obtain
\begin{eqnarray*}
{G}_{13}&=&g^2{\cal C}_F\mu^\epsilon \frac{n^2}{2p^+}
\int\frac{d^{2-\epsilon}l}{(2\pi)^{3-\epsilon}}
\left[\frac{n^+l_T^2}{\sqrt{(p\cdot n)^2-n^2l_T^2}^3}
\ln\frac{\sqrt{(p\cdot n)^2-n^2l_T^2}+p\cdot n}
{\sqrt{(p\cdot n)^2-n^2l_T^2}-p\cdot n}\right.
\nonumber \\
& &\left.-\frac{2n^+p\cdot n}{n^2[(p\cdot n)^2-n^2l_T^2]}
-\frac{2p^+}{n^2(l_T^2+a^2)}\right]\;.
\nonumber
\label{g13}
\end{eqnarray*}

It is obvious that choosing $n^2<0$ avoids extra singularities in the
above integral. The first and second terms have no infrared divergences, 
but the third term does. Hence, we have introduced a small cutoff $a$ 
into the denominator of the third term as in the evaluation of ${K}_{12}$. 
After a tedious calculation, we arrive at
\begin{eqnarray*}
{G}_{13}=-\frac{\alpha_s}{2\pi}{\cal C}_F
\ln\frac{4(p^+)^2\nu^2}{a^2e}\;,
\end{eqnarray*}
which is ultravioletly finite because of the exact cancellation of poles. 
At last, $G_{13}-K_{13}$ is written as
\begin{eqnarray*}
{G}_{13}-{K}_{13}=-\frac{\alpha_s}{2\pi}{\cal C}_F\left[-\frac{2}{\epsilon}
+\ln\frac{(p^+)^2\nu^2 e^{\gamma_E-1}}{\pi\mu^2}\right]\;.
\end{eqnarray*}
Again, the dependence on the cutoff $a$ disappears. In the ${\overline 
{\rm MS}}$ scheme, we have $\delta{G}=-\delta{K}$ and ${G}=
{G}_{13}-{K}_{13}-\delta{G}$ as
\begin{eqnarray*}
{G}=-\frac{\alpha_s}{\pi}{\cal C}_F\left[\ln\frac{p^+\nu}{\mu}
+\ln\frac{2}{\sqrt{e}}\right]\;.
\label{50}
\end{eqnarray*}
Keeping only the first logarithmic term, we derive Eq.~(\ref{oj}).

\vskip 1.0cm

\noindent
3. The function $K'$ in Eq.~(\ref{kj})

Based on the conclusion for $K$ that only the contribution from the term
$n^\mu l^\nu/(n\cdot l)$ needs to be evaluated, we reexpress 
Eq.~(\ref{kj}) as
\begin{eqnarray*}
\delta(p'^2){K}'&=&-ig^2{\cal C}_F\mu^\epsilon n^2\int
\frac{d^{4-\epsilon}l}{(2\pi)^{4-\epsilon}}\frac{1}{(n\cdot l)^2}
\left[\frac{\delta(p'^2)}{l^2}+2\pi i\delta(l^2)\delta((p'-l)^2)\right]
\nonumber \\
& &-\delta(p'^2)\delta K'\;.
\nonumber
\label{kja}
\end{eqnarray*}
Since the first term has been computed, we concentrate on the second 
term. Similarly, the integration over $l^+$ leads to
\begin{eqnarray*}
\delta(p'^2){K}'_{22}=g^2{\cal C}_F\mu^\epsilon n^2\int
\frac{d^{2-\epsilon}l}{(2\pi)^{3-\epsilon}}
\int_{0}^{\infty}d l^-\frac{2l^-}
{(2n^+l^{-2}+n^-l_T^2)^2}\delta\left(p'^2-p'^-\frac{l_T^2}{l^-}\right)\;.
\end{eqnarray*}
The integration over $l^-$ gives
\begin{eqnarray*}
\delta(p'^2){K}'_{22}=g^2{\cal C}_F\mu^\epsilon n^2\int
\frac{d^{2-\epsilon}l}{(2\pi)^{3-\epsilon}}
\frac{2(p'^-)^2 p'^2}{[2n^+(p'^-)^2(l_T^2+a^2)+n^-p'^4]^2}\;,
\end{eqnarray*}
where the small constant $a$ has been inserted for the similar reason.
Because the above integral does not contain ultraviolet and infrared 
singularities, we set $\epsilon=0$ and perform the
integration over $l_T$. We obtain 
\begin{eqnarray*}
\delta(p'^2){K}'_{22}=\frac{\alpha_s}{\pi}{\cal C}_F
\frac{p'^2}{p'^4+4\nu'^2(p'^-)^2a^2}\;.
\end{eqnarray*}

Now we interpret the right-hand side of the above formula as a
distribution function, and extract the on-shell pole at $p'^2=0$ using
\begin{eqnarray*}
\delta(p'^2){K}'_{22}&=&\frac{\alpha_s}{\pi}{\cal C}_F
\delta(p'^2)\int_0^{2(p'^-)^2}d p'^2
\frac{p'^2}{p'^4+4\nu'^2(p'^-)^2a^2}\;,
\nonumber \\
&=&\delta(p'^2)\frac{\alpha_s}{\pi}{\cal C}_F
\ln\frac{(p'^-)}{\nu' a}\;.
\nonumber
\label{kj2}
\end{eqnarray*}
Combining the above expression with $K_{12}$, we derive Eq.~(\ref{okj})
\begin{eqnarray*}
{K}'=-\frac{\alpha_s}{\pi}{\cal C}_F\ln\frac{\nu'\mu}{p'^-}\;.
\end{eqnarray*}
\vskip 1.0cm

\noindent
4. The function $G'$ in Eq.~(\ref{gj})

The calculation of $G'$ from Fig.~10(d) is basically the same as of
$G$. The treatment of the third and fourth diagrams has been explained
in Sec. III. That is, we replace $p'^-$ by $p^+$ and $a$ by 
$\sqrt{p^+ p'^-}$ in $-K'_{22}$.
\vskip 1.0cm

\noindent
5. The anomalous dimension $\lambda_\phi$ in Eq.~(\ref{php})

We evaluate the anomalous dimension $\lambda_\phi$ of the distribution
function $\phi$ in inclusive $B$ meson decays. As stated in Sec. V, 
single soft logarithms are produced from Figs.~6(b) and 6(c). To extract 
the soft structure of these radiative corrections, we apply the eikonal 
approximation to the $b$ quark line \cite{LY1,LY2}. Fig.~6(b), 
corresponding to the self-energy correction to the $b$ quark, gives a 
contribution proportional to
\begin{eqnarray*}
\int\frac{d^4l}{(2\pi)^4}\frac{\gamma_\mu v_\nu}{v\cdot ll^2}N^{\mu\nu}\;,
\end{eqnarray*}
where $\gamma_\mu$ comes from the outer gluon vertex on the $b$ quark line. 
This integral vanishes because the integrand has an odd power of $l$.

The contribution from Fig.~6(c) has been expressed
in Eq.~(\ref{p1}). As explained in Sec. V, only the terms $g^{\mu\nu}$ and
$n^2l^\mu l^\nu/(n\cdot l)^2$ need to be considered, which lead to equal 
results. Hence, we have the bare distribution function 
\begin{eqnarray*}
\phi^{(0)}&=&2g^2{\cal C}_F\mu^\epsilon\int
\frac{d^{4-\epsilon}l}{(2\pi)^{3-\epsilon}}
\frac{n^2}{(n\cdot l)^2}\delta(l^2)e^{i{\bf l}_T\cdot {\bf b}}
\nonumber \\
&=&2g^2{\cal C}_F\mu^\epsilon\int
\frac{d^{2-\epsilon}l}{(2\pi)^{3-\epsilon}}
\frac{1}{l_T^2}e^{i{\bf l}_T\cdot {\bf b}}\;,
\nonumber
\end{eqnarray*}
from which the infrared pole 
\begin{eqnarray*}
\phi^{(0)}_{\rm pole}=\frac{\alpha_s}{\pi}{\cal C}_F\frac{1}{-\epsilon}
\label{62}
\end{eqnarray*}
is extracted. Therefore, the anomalous dimesion is given by
\begin{eqnarray*}
\lambda_\phi=-\frac{\alpha_s}{\pi}{\cal C}_F\;,
\end{eqnarray*}
as exhibited in Eq.~(\ref{php}).

\newpage
\centerline{\large \bf Figure Captions}
\vskip 0.3cm

\noindent
{\bf Fig. 1.} (a) The jet subprocess defined in Eq.~(\ref{j}).
(b) and (c) Lowest-order diagrams of (a).
\vskip 0.3cm

\noindent
{\bf Fig. 2.} The derivative $p^+dJ/dp^+$ in covariant gauge.
\vskip 0.3cm

\noindent
{\bf Fig. 3.} (a) $O(\alpha_s^2)$ examples for the differentiated $J$.
(b) Factorization of the function $K$ at $O(\alpha_s)$.
(c) Factorization of the function $K$ at $O(\alpha_s^2)$.
(d) Factorization of the function $G$ at $O(\alpha_s)$.
\vskip 0.3cm

\noindent
{\bf Fig. 4.} (a) General diagrams for the functions $K$ and $G$.
(b) The $O(\alpha_s)$ function $K$. (c) The $O(\alpha_s)$ function $G$. 
\vskip 0.3cm

\noindent
{\bf Fig. 5.} (a) The derivative $p^+dJ/dp^+$ in axial gauge.
(b) General diagrams for the functions $K$ and $G$.
(c) The $O(\alpha_s)$ function $K$. (d) The $O(\alpha_s)$ function $G$. 
\vskip 0.3cm

\noindent
{\bf Fig. 6.} (a) Lowest-order diagram of deeply inelastic scattering.
(b), (c), (d) and (e) $O(\alpha_s)$ corrections to (a).
\vskip 0.3cm

\noindent
{\bf Fig. 7.} Factorization of Fig.~6(e) for (a) the loop momentum 
parallel to the momentum of the valence quark, for (b) the loop momentum 
parallel to the momentum of the outgoing quark, and for (c) the soft 
loop momentum.
\vskip 0.3cm

\noindent
{\bf Fig. 8.} Factorization of deeply inelastic scattering.
\vskip 0.3cm

\noindent
{\bf Fig. 9.} (a) The derivative $p^+d\phi/dp^+$ in axial gauge.
(b) General diagrams for the functions $K$ and $G$.
(c) The $O(\alpha_s)$ function $K$. (d) The $O(\alpha_s)$ function $G$. 
\vskip 0.3cm

\noindent
{\bf Fig. 10.} (a) The derivative $p'^-dJ/dp'^-$ in axial gauge.
(b) General diagrams for the functions $K'$ and $G'$.
(c) The $O(\alpha_s)$ function $K'$. (d) The $O(\alpha_s)$ function $G'$. 
\vskip 0.3cm

\noindent
{\bf Fig. 11.} (a) Lowest-order diagram of Drell-Yan production.
(b), (c), (d) and (e) $O(\alpha_s)$ corrections to (a).
\vskip 0.3cm

\noindent
{\bf Fig. 12.} Factorization of Drell-Yan production.

\end{document}